\documentclass[12pt]{article}
\usepackage{graphicx}
\usepackage{lscape}
\usepackage{citesort}
\usepackage{amssymb}
\usepackage{appendix}
\usepackage{multirow}

\begin{document}
\def\qq{\langle \bar q q \rangle}
\def\uu{\langle \bar u u \rangle}
\def\dd{\langle \bar d d \rangle}
\def\sp{\langle \bar s s \rangle}
\def\GG{\langle g_s^2 G^2 \rangle}
\def\Tr{\mbox{Tr}}
\def\figt#1#2#3{
        \begin{figure}
        $\left. \right.$
        \vspace*{-2cm}
        \begin{center}
        \includegraphics[width=10cm]{#1}
        \end{center}
        \vspace*{-0.2cm}
        \caption{#3}
        \label{#2}
        \end{figure}
    }

\def\figb#1#2#3{
        \begin{figure}
        $\left. \right.$
        \vspace*{-1cm}
        \begin{center}
        \includegraphics[width=10cm]{#1}
        \end{center}
        \vspace*{-0.2cm}
        \caption{#3}
        \label{#2}
        \end{figure}
                }

\def\ds{\displaystyle}
\def\beq{\begin{equation}}
\def\eeq{\end{equation}}
\def\bea{\begin{eqnarray}}
\def\eea{\end{eqnarray}}
\def\beeq{\begin{eqnarray}}
\def\eeeq{\end{eqnarray}}
\def\ve{\vert}
\def\vel{\left|}
\def\ver{\right|}
\def\nnb{\nonumber}
\def\ga{\left(}
\def\dr{\right)}
\def\aga{\left\{}
\def\adr{\right\}}
\def\lla{\left<}
\def\rra{\right>}
\def\rar{\rightarrow}
\def\lrar{\leftrightarrow}
\def\nnb{\nonumber}
\def\la{\langle}
\def\ra{\rangle}
\def\ba{\begin{array}}
\def\ea{\end{array}}
\def\tr{\mbox{Tr}}
\def\ssp{{\Sigma^{*+}}}
\def\sso{{\Sigma^{*0}}}
\def\ssm{{\Sigma^{*-}}}
\def\xis0{{\Xi^{*0}}}
\def\xism{{\Xi^{*-}}}
\def\qs{\la \bar s s \ra}
\def\qu{\la \bar u u \ra}
\def\qd{\la \bar d d \ra}
\def\qq{\la \bar q q \ra}
\def\gGgG{\la g^2 G^2 \ra}
\def\q{\gamma_5 \not\!q}
\def\x{\gamma_5 \not\!x}
\def\g5{\gamma_5}
\def\sb{S_Q^{cf}}
\def\sd{S_d^{be}}
\def\su{S_u^{ad}}
\def\sbp{{S}_Q^{'cf}}
\def\sdp{{S}_d^{'be}}
\def\sup{{S}_u^{'ad}}
\def\ssp{{S}_s^{'??}}

\def\sig{\sigma_{\mu \nu} \gamma_5 p^\mu q^\nu}
\def\fo{f_0(\frac{s_0}{M^2})}
\def\ffi{f_1(\frac{s_0}{M^2})}
\def\fii{f_2(\frac{s_0}{M^2})}
\def\O{{\cal O}}
\def\sl{{\Sigma^0 \Lambda}}
\def\es{\!\!\! &=& \!\!\!}
\def\ap{\!\!\! &\approx& \!\!\!}
\def\md{\!\!\!\! &\mid& \!\!\!\!}
\def\ar{&+& \!\!\!}
\def\ek{&-& \!\!\!}
\def\kek{\!\!\!&-& \!\!\!}
\def\cp{&\times& \!\!\!}
\def\se{\!\!\! &\simeq& \!\!\!}
\def\eqv{&\equiv& \!\!\!}
\def\kpm{&\pm& \!\!\!}
\def\kmp{&\mp& \!\!\!}
\def\mcdot{\!\cdot\!}
\def\erar{&\rightarrow&}
\def\olra{\stackrel{\leftrightarrow}}
\def\ola{\stackrel{\leftarrow}}
\def\ora{\stackrel{\rightarrow}}

\def\simlt{\stackrel{<}{{}_\sim}}
\def\simgt{\stackrel{>}{{}_\sim}}


\title{
         {\Large
                 {\bf
                      Thermal properties of $D_2^*(2460)$ and $D_{s2}^*(2573)$ tensor mesons using QCD sum rules
                 }
         }
      }

\author{{\small K. Azizi$^{\dag}$,  H. Sundu$^{*}$, A. T\"urkan$^{*}$ and E. Veli Veliev$^{*}$} \\
{\small $^{\dag}$ Department of Physics,
Do\u gu\c s University, Ac{\i}badem-Kad{\i}k\"oy, 34722 Istanbul, Turkey}\\
{\small$^{*}$Department of Physics, Kocaeli University, 41380 Izmit,
Turkey}\\
}
\date{}

\begin{titlepage}
\maketitle
\thispagestyle{empty}

\begin{abstract}
We investigate the masses and decay constants of the heavy-light
 $D_2^*(2460)$ and $D_{s2}^*(2573)$ tensor mesons in the framework
of thermal QCD sum rules. Taking into account the additional operators
arising at finite temperature, we evaluate the Wilson expansion for
the two-point correlation function associated with these
mesons. We observe that the values of the masses and decay constants decrease
considerably at near to the critical temperature.  The decay
constants attain roughly to $25\%$ of their values in vacuum,
while the masses decrease about $39\%$ and $37\%$ in $D_2^*$ and
$D_{s2}^*$ channels, respectively. 
\end{abstract}

~~~PACS number(s): 11.55.Hx, 11.10.Wx, 14.40.Lb
\end{titlepage}

\section{Introduction}
During the last few decades, many tensor mesons have been observed by different
experiments
\cite{W.D.Apel,R.S.Longacre,M.Doser,Y.Kubota,P.Avery,T.Bergfeld,M.Ablikim}.
The investigation  of these particles is one of the most
interesting problems in hadron physics both theoretically and experimentally.   In the literature, there
are few theoretical works devoted to the analysis of  the properties of the
tensor mesons compared to the scalar, pseudoscalar, vector and axial-vector mesons. The study of parameters of tensor mesons and their
comparison with the experimental results can give useful information
on their nature and internal structure. Moreover,  the investigation  of these particles  can  be
useful for understanding the non-perturbative dynamics as well as the vacuum
structure of QCD. 

The observation of charmed  $D_2^*(2460)$ and $D_{s2}^*(2573)$
states both with quantum numbers $J^{P}=2^{+}$, were reported 
 twenty years ago \cite{Y.Kubota,P.Avery,T.Bergfeld} and confirmed by the LHCb collaboration in 2011 \cite{lhcb}.
The properties of these mesons at zero temperature have been
recently studied in \cite{hayriye,jale}. In this article, we investigate the  thermal properties of
these  particles, which can be used  in analysis of  the results of the heavy ion
collisions held at different experiments.

The study of parameters of  mesons  at finite temperature requires some
thermal non-perturbative approaches. One of the most attractive and applicable tools in this respect is the
thermal QCD sum rules firstly suggested for investigation
of hadronic parameters in vacuum  \cite{Shifman} and  later was extended
to finite temperature and density \cite{Bochkarev}. This
extension was based on some basic assumptions so that the Wilson
expansion and the quark-hadron duality approximation remain valid, but the
vacuum condensates are replaced by their thermal expectation
values. 
 At finite temperature, the 
Lorentz invariance is broken by the choice of a preferred frame of
reference and some new operators appear in the Wilson expansion
\cite{E.V. Shuryak,T. Hatsuda,S. Mallik,K. Mukherjee}. To
restore the Lorentz invariance in thermal field theory, the four-vector
velocity of the medium is introduced. Making use of this
velocity and the fermionic and gluonic parts of the energy-momentum tensor,  a new set of four dimensional  operators are constructed. The thermal QCD sum
rule method has been widely used to investigate the medium
properties of the light-light \cite{A. Nyfeler,Veliev}, the heavy-light
\cite{Dominguez,Loewe,Veliev2} and the heavy-heavy
\cite{Klingl,Morita,Lee,Veliev3,arzu,nurcan} systems mainly in recent years.

In the present work, in particular, we investigate the masses and decay constants of the 
$D_2^*$ and $D_{s2}^*$ tensor mesons in the framework of thermal
QCD sum rules method. Taking into account the additional operators
coming up at finite temperature, we calculate the thermal two
point correlation function and  obtain the spectral densities
in one loop approximation. In order to perform the numerical analysis,
we use the fermionic part of the energy density  obtained both from
lattice QCD \cite{M.Cheng,D.E.Miller} and Chiral perturbation
theory \cite{P.Gerber}. We also use the temperature dependent
continuum threshold \cite{Dominguez} and  investigate the
 sensitivity of the masses and decay constants to the temperature. 

The paper is organized as follows. In section 2 we evaluate the Wilson
expansion for the two-point correlation function and  derive the
thermal QCD sum rules for the masses and decay constants of the $D_{2}^*$ and $D_{s2}^*$ tensor states. In
section 3 we present our numerical calculations and discuss the obtained results.
\section{Theoretical framework}
To calculate the masses and decay constants of the $D_2^*$ and
$D_{s2}^*$ tensor mesons in the framework of the thermal QCD sum
rules, we start with  the following thermal correlation
function:
\begin{eqnarray}\label{correl.func.101}
\Pi _{\mu\nu,\alpha\beta}(q,T)=i\int
d^{4}xe^{iq\cdot(x-y)}{\langle} {\cal T}[j _{\mu\nu}(x) \bar
j_{\alpha\beta}(y)]{\rangle}_{|_{y=0}},
\end{eqnarray}
where $j_{\mu\nu}$ is the interpolating current of the tensor
mesons, $T$ is temperature and $\cal T$ indicates the time ordering operator. As the interpolating current of the tensor mesons contains derivatives with respect to the space-time, 
after applying derivatives with respect to $y$ we will set $y=0$. The
thermal average of any operator $A$ in thermal equilibrium is defined as
${\langle}A{\rangle}=\frac{Tr(e^{-\beta H} A)}{Tr( e^{-\beta
H})}$, where $H$ is the QCD Hamiltonian and $\beta=1/T$. 

The interpolating
current $j_{\mu\nu}$ for tensor mesons is written as

\begin{eqnarray}\label{tensorcurrent}
j _{\mu\nu}(x)=\frac{i}{2}\left[\bar q(x) \gamma_{\mu} \olra{\cal
D}_{\nu}(x) c(x)+\bar q(x) \gamma_{\nu}  \olra{\cal D}_{\mu}(x)
c(x)\right],
\end{eqnarray}
where $q$ is $u$ ($s$) quark for $D_2^*$ ($D_{s2}^*$)  and $ \olra{\cal
D}_{\mu}(x)$ denotes the four-derivative with respect to $x$
acting on the left and right, simultaneously. It is given as
\begin{eqnarray}\label{derivative}
\olra{\cal D}_{\mu}(x)=\frac{1}{2}\left[\ora{\cal D}_{\mu}(x)-
\ola{\cal D}_{\mu}(x)\right],
\end{eqnarray}
where
\begin{eqnarray}\label{derivative2}
\overrightarrow{{\cal
D}}_{\mu}(x)=\overrightarrow{\partial}_{\mu}(x)-i
\frac{g}{2}\lambda^aA^a_\mu(x),\nonumber\\
\overleftarrow{{\cal
D}}_{\mu}(x)=\overleftarrow{\partial}_{\mu}(x)+
i\frac{g}{2}\lambda^aA^a_\mu(x).
\end{eqnarray}
Here,  $\lambda^a$ ($a=1,~2.....8$) are the Gell-Mann matrices and $A^a_\mu(x)$ are
the external  gluon fields.

According to the basic idea in the QCD sum rule method, the aforementioned thermal
correlation function can be calculated in two different ways: first, in terms of QCD degrees of freedom called
theoretical or QCD side, and the second, in terms of hadronic parameters called the  physical or phenomenological
side. The correlation function in QCD side is
calculated using the operator product expansion (OPE), where the short  and long distance effects (see figure 1)  are separated.  The thermal QCD sum rules for the physical observables
such as the masses and decay constants  are obtained equating
the coefficients of the same structure
from both sides of the correlation function  through a dispersion relation. Finally, the Borel transformation  and continuum subtraction are performed in order to suppress the contributions
of the higher states and continuum.
\begin{figure}[h!]
\begin{center}
\includegraphics[totalheight=4cm,width=10cm]{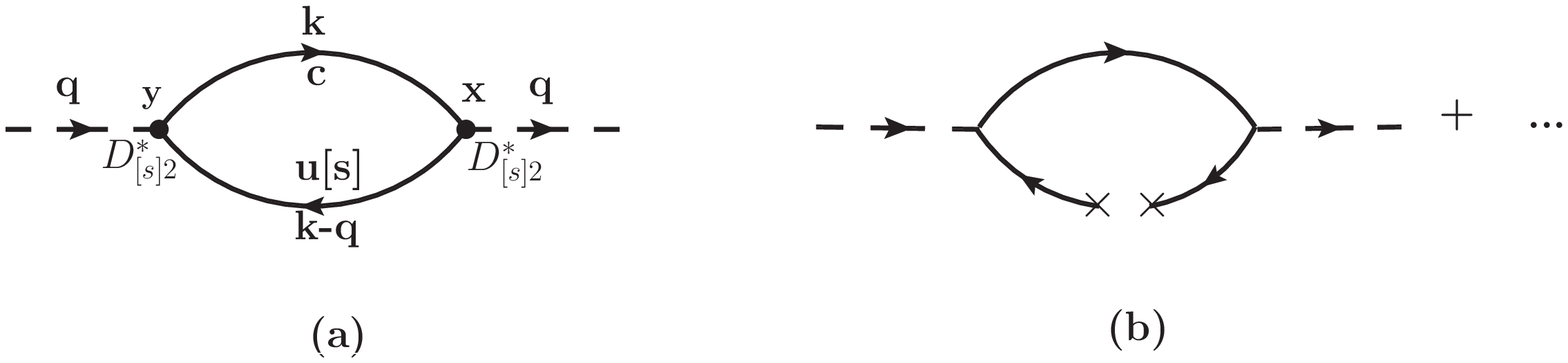}
\end{center}
\caption{ (a) Bare loop diagram (short-distance or perturbative contribution); (b) quark condensate diagram (the lowest dimension long-distance or non-perturbative contribution).}
\label{Diagrams1}
\end{figure}

\subsection{Correlation function in  QCD representation}

As previously mentioned, the correlation function in QCD side is evaluated via OPE in deep Euclidean region where the perturbative and non-perturbative contributions
are separated. The perturbative part is calculated via perturbation theory using spectral representation, while the
non-perturbative contributions are represented in terms of the
thermal expectation values of the quark and gluon condensates as
well as thermal average of the energy density. 
Putting the expression of the interpolating current and
covariant four derivatives into correlation function in Eq. (\ref{correl.func.101}) and
applying the Wick's theorem, we get
\begin{eqnarray}\label{correl.func.2}
\Pi _{\mu\nu,\alpha\beta}&=&-\frac{i}{16}\int d^{4}xe^{iq.(x-y)}
\Bigg\{Tr\Big[\ora{\cal D}_{\beta}(y)S_q(y-x)\gamma_\mu\ora{\cal
D}_{\nu}(x)S_c(x-y)\gamma_\alpha
\nonumber\\&-&S_q(y-x)\gamma_\mu\ora{\cal D}_{\nu}(x)\ora{\cal
D}_{\beta}(y)S_c(x-y)\gamma_\alpha-\ora{\cal
D}_{\beta}(y)\ora{\cal D}_{\nu}(x)S_q(y-x)\gamma_\mu
S_c(x-y)\gamma_\alpha\nonumber\\&+&\ora{\cal
D}_{\nu}(x)S_q(y-x)\gamma_\mu\ora{\cal
D}_{\beta}(y)S_c(x-y)\gamma_\alpha\Big]+\left[\beta\leftrightarrow\alpha\right]
+\left[\nu\leftrightarrow\mu\right]\nonumber\\&+&\left[\beta\leftrightarrow\alpha,
\nu\leftrightarrow\mu\right]\Bigg\},
\end{eqnarray}
where we kept only the full contracted terms. The normally ordered terms also give non-perturbative contributions which we  take into account  in the expressions of the propagators.
The expressions for the heavy quark propagator $S_c(x-y)$ and the light quark propagator
$S_q(x-y)$   in coordinate space, up to the terms considered in the present work,  are given as
\begin{eqnarray}\label{heavypropagator}
S_{c}^{ij}(x-y)=\frac{i}{(2\pi)^4}\int d^4k e^{-ik \cdot (x-y)}
\left\{ \frac{\!\not\!{k}+m_c}{k^2-m_c^2}\delta_{ij}
 +\cdots\right\} \, ,
\end{eqnarray}
and
\begin{eqnarray}
S_{q}^{ij}(x-y)&=& i\frac{\!\not\!{x}-\!\not\!{y}}{
2\pi^2(x-y)^4}\delta_{ij}
-\frac{m_q}{4\pi^2(x-y)^2}\delta_{ij}-\frac{\langle
\bar{q}q\rangle}{12}\delta_{ij}+\frac{i}{3}\Big[(\!\not\!{x}-\!\not\!{y})\Big(\frac{m_q}{16}\langle
\bar{q}q\rangle-\frac{1}{12}\langle u\Theta^{f}u\rangle\Big)
\nonumber\\
&+&\frac{1}{3}\Big(u.(x-y)\!\not\!{u}\langle
u\Theta^{f}u\rangle\Big)\Big]\delta_{ij}+....,
\end{eqnarray}
where $\Theta^{f}_{\mu\nu}$ is the fermionic part of the energy momentum tensor and  $u_{\mu}$ is the four-velocity of the heat
bath. In the rest frame of the heat bath, $u_{\mu}=(1,0,0,0)$ and
$u^2=1$. 
Note that in our calculations  we ignore the two-gluon condensate terms   because of
their small contributions (see also  \cite{aliev,Kazim1,ZGWang}).

The  next step is to use the expressions of the propagators in Eq.
(\ref{correl.func.2}) and apply the derivatives with respect to
$x$ and $y$. After setting $y = 0$, we get
\begin{eqnarray}\label{correl.func.3}
\Pi _{\mu\nu,\alpha\beta}&=&\frac{1}{16}
 \int\frac{d^4k}{(2\pi)^4}\frac{1}{k^2-m_c^2}\int
d^{4}xe^{i(q-k)\cdot
x}\Bigg\{\Big[Tr\Gamma_{\mu\nu,\alpha\beta}\Big]+
\Big[\beta\leftrightarrow\alpha\Big]+\Big[\nu\leftrightarrow\mu\Big]\nonumber\\&+&
\Big[\beta\leftrightarrow\alpha,\nu\leftrightarrow\mu\Big]\Bigg\},
\end{eqnarray}
where $\Gamma_{\mu\nu,\alpha\beta}$ is given by
\begin{eqnarray}\label{fonk}
\Gamma_{\mu\nu,\alpha\beta}&=&
k_{\nu}k_{\beta}\Big[\frac{i\!\not\!{x}}{2\pi^2 x^4}
+\frac{m_q}{4\pi^2x^2}+\Big(\frac{1}{12}+\frac{im_q
\!\not\!{x}}{48}\Big)\langle
\bar{q}q\rangle-\Big(\frac{i\!\not\!{x}}{36}-\frac{i}{9}u\cdot x
\!\not\!{u}\Big)\langle
u\Theta^{f}u\rangle\Big]\gamma_{\mu}(\!\not\!{k}+m_c)\gamma_{\alpha}
\nonumber\\
&-&ik_{\nu} \Big[\frac{i}{2
\pi^2}\Big(\frac{\gamma_{\beta}}{x^4}-\frac{4x_{\beta}\!\not\!{x}}{x^6}\Big)-\frac{m_q
x_{\beta}}{2\pi^2
x^4}+\frac{i\gamma_{\beta}m_q}{48}\langle\bar{q}q\rangle-\Big(\frac{i\gamma_{\beta}}{36}-\frac{i}{9}u_{\beta}\!\not\!{u}\Big)\langle
u\Theta^{f}u\rangle\Big]\gamma_{\mu}(\!\not\!{k}+m_c)\gamma_{\alpha}
\nonumber\\
&+&ik_{\beta}\Big[\frac{i}{2
\pi^2}\Big(\frac{4x_{\nu}\!\not\!{x}}{x^6}
-\frac{\gamma_{\nu}}{x^4}\Big)+\frac{m_q x_{\nu}}{2 \pi^2
x^4}-\frac{im_q\gamma_{\nu}}{48}\langle\bar{q}q\rangle+\Big(\frac{i\gamma_{\nu}}{36}-\frac{i}{9}u_{\nu}\!\not\!{u}\Big)\langle
u\Theta^{f}u\rangle\Big]\gamma_{\mu}(\!\not\!{k}+m_c)\gamma_{\alpha}
\nonumber\\
&+&\Big[\frac{8i}{2\pi^2}\Big(\frac{ x_{\beta}
x_{\nu}\!\not\!{x}}{x^{8}}+\frac{1}{2x^{6}}
\Big(\delta^{\nu}_{\beta}\!\not\!{x}-\gamma_{\nu}
x_{\beta}+\gamma_{\beta} x_{\nu}\Big) +\frac{ \gamma_{\nu}
x_{\beta}}{x^6}-\frac{4
x_{\beta}x_{\nu}\!\not\!{x}}{x^8}\Big)+\frac{\delta^{\nu}_{\beta}m_q}{2\pi^2
x^4}+\frac{2 m_q x_{\nu}x_{\beta}}{\pi^2 x^6}\Big]
\nonumber\\
&\times& \gamma_{\mu} (\!\not\!{k}+m_c)\gamma_{\alpha}+
\left[\beta\leftrightarrow\alpha\right]+\left[\nu\leftrightarrow\mu\right]+
\left[\beta\leftrightarrow\alpha,\nu\leftrightarrow\mu\right].
\end{eqnarray}
%

%
\subsection{Correlation function in  phenomenological representation}

To calculate the phenomenological side of the correlation
function, a complete set of physical intermediate states having the
same quantum numbers as the interpolating current is inserted into
Eq. (\ref{correl.func.101}). After performing integral over $x$
and putting $y = 0$, we obtain 
\begin{eqnarray}\label{phen1}
\Pi _{\mu\nu,\alpha\beta}^{D_2^*(D_{s2}^*)}=\frac{{\langle}0\mid  j _{\mu\nu}(0)
\mid D_2^*(D_{s2}^*)\rangle \langle D_2^*(D_{s2}^*)\mid \bar
j_{\alpha\beta}(0)\mid
 0\rangle}{m_{D_2^*(D_{s2}^*)}^2-q^2}
&+& \cdots,
\end{eqnarray}
where  dots indicate the contributions of the higher states
and continuum. The matrix element $\langle 0 \mid
j_{\mu\nu}(0)\mid D_2^*(D_{s2}^*)\rangle$  can be written in terms of the decay constant
$f_{D_2^*(D_{s2}^*)}$ as
\begin{eqnarray}\label{lep}
\langle 0 \mid j_{\mu\nu}(0)\mid
D_2^*(D_{s2}^*)\rangle=f_{D_2^*(D_{s2}^*)}
m_{D_2^*(D_{s2}^*)}^3\varepsilon_{\mu\nu}^{(\lambda)}.
\end{eqnarray}
where 
$\varepsilon_{\mu\nu}^{(\lambda)}$ is the polarization tensor. We use the 
summation over polarization tensors as
\begin{eqnarray}\label{polarizationt1}
\sum_{\lambda}\varepsilon_{\mu\nu}^{(\lambda)}\varepsilon_{\alpha\beta}^{*(\lambda)}=\frac{1}{2}\eta_{\mu\alpha}\eta_{\nu\beta}+
\frac{1}{2}\eta_{\mu\beta}\eta_{\nu\alpha}
-\frac{1}{3}\eta_{\mu\nu}\eta_{\alpha\beta},
\end{eqnarray}
where
\begin{eqnarray}\label{polarizationt2}
\eta_{\mu\nu}=-g_{\mu\nu}+\frac{q_\mu
q_\nu}{m_{D_2^*(D_{s2}^*)}^2}.
\end{eqnarray}
Using the above expressions in Eq. (\ref{phen1}), the final
representation of the physical side is obtained as
\begin{eqnarray}\label{phen2}
\Pi _{\mu\nu,\alpha\beta}^{D_2^*(D_{s2}^*)}=\frac{f_{D_2^*(D_{s2}^*)}^2
m_{D_2^*(D_{s2}^*)}^4} {m_{D_2^*(D_{s2}^*)}^2-q^2}
\left\{-\frac{1}{2}q_{\mu}q_{\alpha}g_{\nu\beta}\right\}+
\mbox{other structures}+...,
\end{eqnarray}
where the explicitly written  structure is used  to extract the QCD  sum rules for the physical quantities under consideration.

\subsection{Thermal QCD sum rules for physical observables}
To obtain the QCD sum rules for the masses and decay constants we need to calculate the perturbative and non-perturbative parts of the correlation function in momentum space then match the coefficients
of the selected structure from both phenomenological and QCD sides. 
 For this aim we write the perturbative part of the  correlation function in QCD side  in terms of a dispersion integral 
as
\begin{eqnarray}\label{QCD Side}
\Pi^{pert}(q,T) =\int \frac{ds \rho(s)}{s-q^2},
\end{eqnarray}
where $\rho(s)$  is the spectral density and it is obtained via the
imaginary part of the perturbative part of the thermal correlator
\begin{eqnarray}\label{QCD Side}
\rho(s)=\frac{1}{\pi}Im[\Pi^{pert}(s)].
\end{eqnarray}
Following the procedures represented in 
 \cite{hayriye,jale}, and after lengthy calculations, we obtain the spectral densities corresponding to the
tensor  $D_2^*$ and $D_{s2}^*$ states as

\begin{eqnarray}\label{spectral densty}
\rho_{D_{2}^*}(s)=-\frac{N_c}{640\pi^2s^4}(m_{c}^2-s)^2(8m_{c}^6-4m_{c}^4s-m_{c}^2s^2+2s^3),
\end{eqnarray}
and
\begin{eqnarray}\label{spectral densty}
\rho_{D_{s2}^*}(s)&=&-\frac{N_c}{1920\pi^2s^4}(m_{c}^2-s)(24m_{c}^8-36m_{c}^6s+40m_{c}^5m_s
s+9m_{c}^4s^2-20m_{c}^3m_ss^2
\nonumber\\
&+&9m_{c}^2s^3+10m_{c}m_ss^3-6s^4),
\end{eqnarray}
where $N_c=3$ is the number of colors.
From a similar way we calculate the non-perturbative contributions (see also \cite{hayriye,jale}).

The final task is to match the phenomenological and QCD sides of the
correlation function in momentum space and apply Borel transformation with respect to
$Q^2=-q^2$. After continuum subtraction we get
\begin{eqnarray}\label{rhomatching}
f_{D_2^*(D_{s2}^*)}^2(T)m_{D_2^*(D_{s2}^*)}^4(T)
e^{-m_{D_2^*(D_{s2}^*)}^2(T)/M^2} =\int_{(m_q+m_c)^2}^{s_0(T)} ds
\rho_{D_2^*(D_{s2}^*)}(s)e^{-s/M^2}+\hat{B}\Pi_{D_2^*(D_{s2}^*)}^{non-pert},
\end{eqnarray}
where $s_0(T)$  is the temperature-dependent continuum threshold and
$M^2$ is the Borel mass parameter. The function $\hat{B}\Pi^{non-pert}$
shows the non-perturbative part of the QCD side in the Borel transformed
scheme. It is given in $D_2^*$ and $D_{s2}^*$ channels  as
\begin{eqnarray}\label{borelnonpertDs}
\hat{B}\Pi_{D_2^*}^{non-pert} =-\frac{m_{c}\langle
\bar{u}u\rangle}{48} e^{-m_{c}^2/M^2}+\frac{\langle
u\Theta^{f}u\rangle}{72}
e^{-m_{c}^2/M^2}-\frac{(-m_{c}^2+M^2)\langle
u\Theta^{f}u\rangle}{24M^2}e^{-m_{c}^2/M^2},
\end{eqnarray}
and
\begin{eqnarray}\label{borelnonpertDs2}
\hat{B}\Pi_{D_{s2}^*}^{non-pert} &=&-\frac{m_c\langle
\bar{s}s\rangle}{48} e^{-m_{c}^2/M^2}+\frac{m_s\langle
\bar{s}s\rangle}{96} e^{-m_{c}^2/M^2}-\frac{m_s
(-m_{c}^2-M^2)\langle \bar{s}s\rangle}{96M^2}e^{-m_{c}^2/M^2}
\nonumber\\
&+&\frac{\langle u\Theta^{f}u\rangle}{72}
e^{-m_{c}^2/M^2}-\frac{(-m_{c}^2+M^2)\langle
u\Theta^{f}u\rangle}{24M^2}e^{-m_{c}^2/M^2}.
\end{eqnarray}
The temperature-dependent masses of the states under consideration are found as
\begin{eqnarray}\label{rhomass}
m_{D_2^*(D_{s2}^*)}^2(T) =\frac{\int_{(m_q+m_c)^2}^{s_0(T)}ds
\rho_{D_2^*(D_{s2}^*)}(s)~s~e^{-s/M^2}+{\cal \psi}_{D_2^*(D_{s2}^*)}^{non-pert}(M^2,T)}
{\int_{(m_q+m_c)^2}^{s_0(T)}ds \rho_{D_2^*(D_{s2}^*)}(s)
e^{-s/M^2}+\hat{B}\Pi_{D_2^*(D_{s2}^*)}^{non-pert}},
\end{eqnarray}
where ${\cal \psi}_{D_2^*(D_{s2}^*)}^{nonpert}(M^2,T)$ is given by
\begin{eqnarray}\label{turevnonpert}
{\cal \psi}_{D_2^*(D_{s2}^*)}^{non-pert}(M^2,T)=M^4
\frac{d}{dM^2}\hat{B}\Pi_{D_2^*(D_{s2}^*)}^{non-pert}.
\end{eqnarray}

\section{Numerical results and discussion}

In this section we present our numerical results on the physical quantities under consideration and discuss their 
sensitivity to the temperature. We also compare the obtained numerical values at $T=0$  with the existing experimental data
\cite{K.Nakamura} and  those obtained from vacuum sum rules \cite{hayriye,jale}. 
For this aim, we use some input  parameters as $m_{s}=0.12$ GeV,  $m_{c}=(1.27^{+0.07}_{-0.09})$ GeV \cite{K.Nakamura},  $\langle0|\overline{u}u|0\rangle=-(0.24\pm0.01)^3~$GeV$^3$ 
\cite{B.L.Ioffe} and $\langle0|\overline{s}s|0\rangle=0.8\langle0|\overline{u}u|0\rangle$ \cite{S. Narison}.

To proceed further, we use the fermionic part of the energy density
 obtained from both lattice QCD \cite{M.Cheng,D.E.Miller} and
Chiral perturbation theory \cite{P.Gerber}. The thermal average of
the energy density obtained using the lattice QCD is expressed as
\begin{eqnarray}\label{turevnonpert}
\langle\Theta\rangle=2\langle\Theta^{f}\rangle=6\times10^{-6}exp\Big[80(T-0.1)\Big],
\end{eqnarray}
where  $T$ is in the units of $GeV$ and this
parametrization is valid only in the region $0.1 GeV\leq T \leq
0.175 GeV$. Here we should mention that the total energy density
has been calculated for $T\geq 0$ in Chiral perturbation theory,
while it is available  only  for $T\geq 100 MeV$ in
lattice QCD \cite{M.Cheng,D.E.Miller}. In the limit of 
low temperature Chiral perturbation, the thermal average of
the energy density is written as \cite{P.Gerber}
\begin{eqnarray}\label{tetagchiral}
\langle \Theta\rangle= \langle \Theta^{\mu}_{\mu}\rangle +3~p,
\end{eqnarray}
where $\langle \Theta^{\mu}_{\mu}\rangle$ is trace of the total
energy momentum tensor and $p$ is pressure. These quantities are
given by

\begin{eqnarray}\label{tetamumu}
\langle
\Theta^{\mu}_{\mu}\rangle&=&\frac{\pi^2}{270}\frac{T^{8}}{F_{\pi}^{4}}
\ln \Big(\frac{\Lambda_{p}}{T}\Big),
\end{eqnarray}
and
\begin{eqnarray}\label{pressure}
p&=&
3T\Big(\frac{m_{\pi}~T}{2~\pi}\Big)^{\frac{3}{2}}\Big(1+\frac{15~T}{8~m_{\pi}}+\frac{105~T^{2}}{128~
m_{\pi}^{2}}\Big)exp\Big(-\frac{m_{\pi}}{T}\Big).
\end{eqnarray}
In further analysis, we also use the light quark condensate at finite
temperature. The temperature-dependent quark condensate
obtained in Chiral perturbation theory \cite{J. Gasser,P.Gerber}
can be written in a good approximation as
\begin{eqnarray}\label{qbarq}
\langle\bar{q}q\rangle=\langle
0|\bar{q}q|0\rangle\Big[1-0.4\Big(\frac{T}{T_c}\Big)^4-0.6\Big(\frac{T}{T_c}\Big)^8\Big].
\end{eqnarray}
where $T_c=0.175$ GeV  \cite{cambridge} is the critical temperature. 
The continuum threshold also depends on the temperature and it is given in terms of the quark condensate
 by \cite{Dominguez}
\begin{equation}\label{eqn16}
s_{0}(T)=s_{0}\frac{\langle\bar{q}q\rangle}{\langle0|\bar{q}q|0\rangle}\Big{(}1-\frac{(m_{c}+m_{q})^{2}}{s_{0}}\Big{)}+(m_{c}+m_{q})^{2},  \\
\end{equation}
where $s_{0}$  in
the right hand side is the hadronic threshold at zero
temperature, i.e., $s_{0}=s(T=0)$. The continuum threshold is not totally arbitrary but it depends on the energy of the first excited state with the same quantum numbers as the chosen interpolating
current. According to the standard procedure in QCD sum rule approach   the working region for this parameter is chosen such that the variations of the results with respect to this parameter in
 the chosen Borel window are weak.
 We choose the intervals $s_{0}=(7.8\pm0.3)~GeV^2$ and
$s_{0}=(9.1\pm0.3)~GeV^2$ for the continuum threshold in the $D_2^*$
and $D_{s2}^*$ channels, respectively. Our analysis show that the dependences of the results on this parameter are very weak in these intervals.

From the sum rules for the physical quantities in the previous section it is clear that they also include an auxiliary Borel parameter
$M^2$ which we shall also find its working region. The working region for the Borel
parameter is found such that not only the contributions of the higher states and continuum are suppressed but also the perturbative part exceeds the non-perturbative contributions and the  contributions 
of the higher dimensional operators 
are small, i.e., the OPE converges. As a result we obtain the interval $ 3~ GeV^2 \leq M^2 \leq 6~ GeV^2 $ for the working region of Borel mass.  Our numerical results show that the contribution of the higher states and continuum
are approximately $10\%$ of the total dispersion integral in the selected regions for the auxiliary parameters. To see how  the results depend on the Borel mass parameter, we plot the dependences 
of the masses and decay constants of the mesons under consideration versus $M^2$ for different values of the continuum threshold at $T=0$ in figures 2 and 3.
 \begin{figure}[h!]
\begin{center}
\includegraphics[width=8cm]{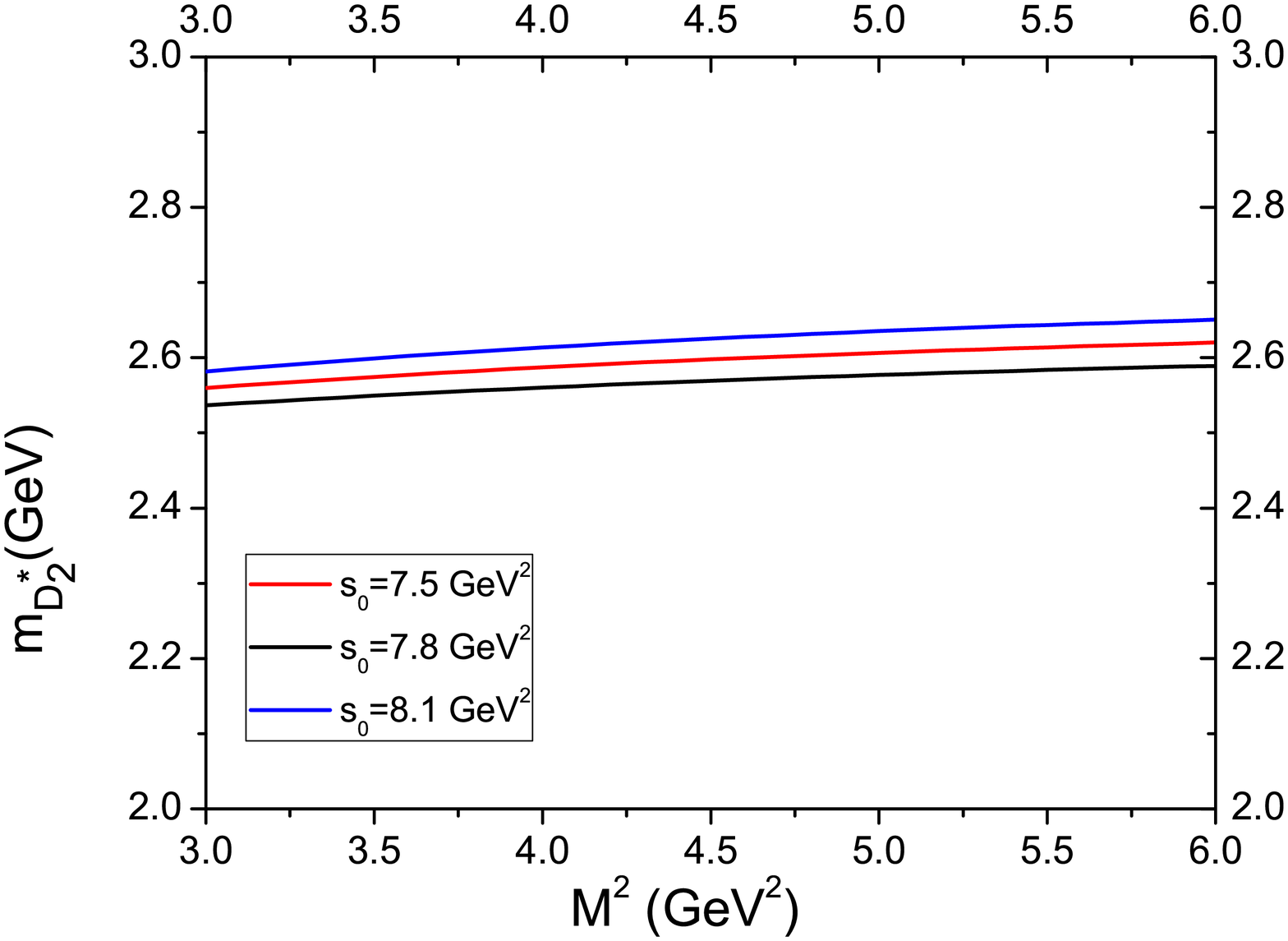}
\includegraphics[width=8cm]{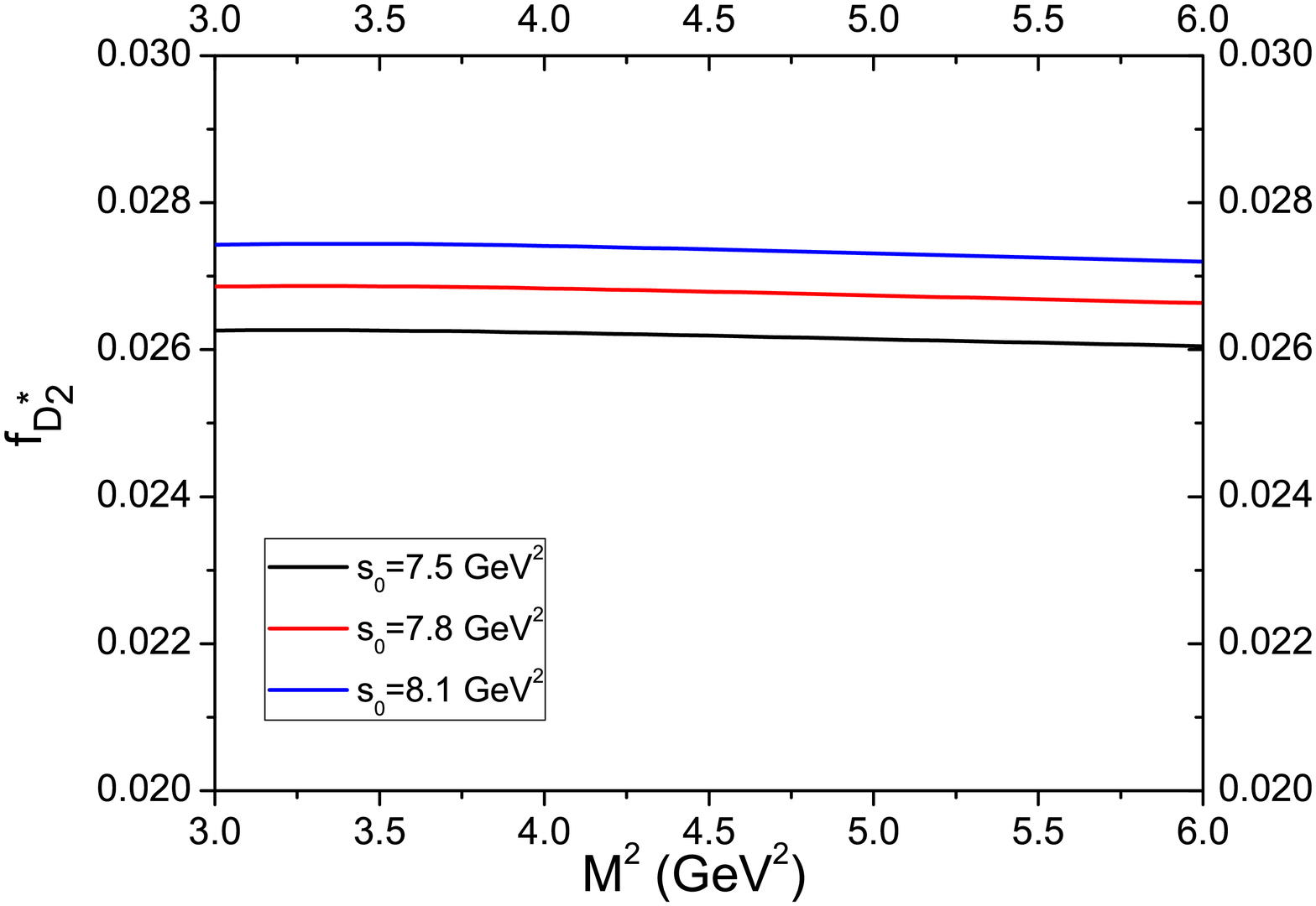}
\end{center}
\caption{Variations of the mass and decay constant
of the $D_2^*(2460)$ meson with respect to $M^2$  at fixed values of the continuum threshold and at $T=0$.} \label{Diagrams1}
\end{figure}
\begin{figure}[h!]
\begin{center}
\includegraphics[width=8cm]{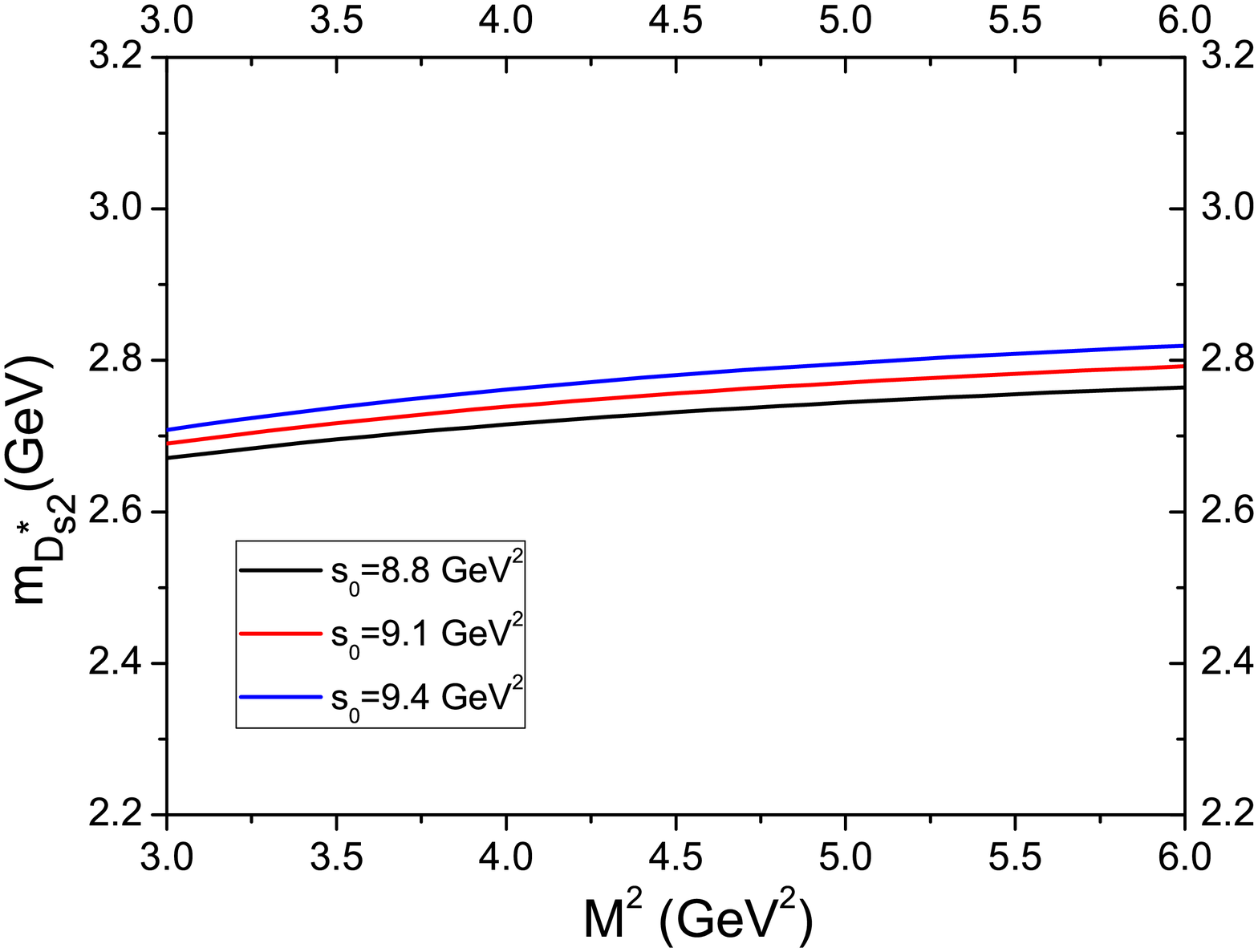}
\includegraphics[width=8cm]{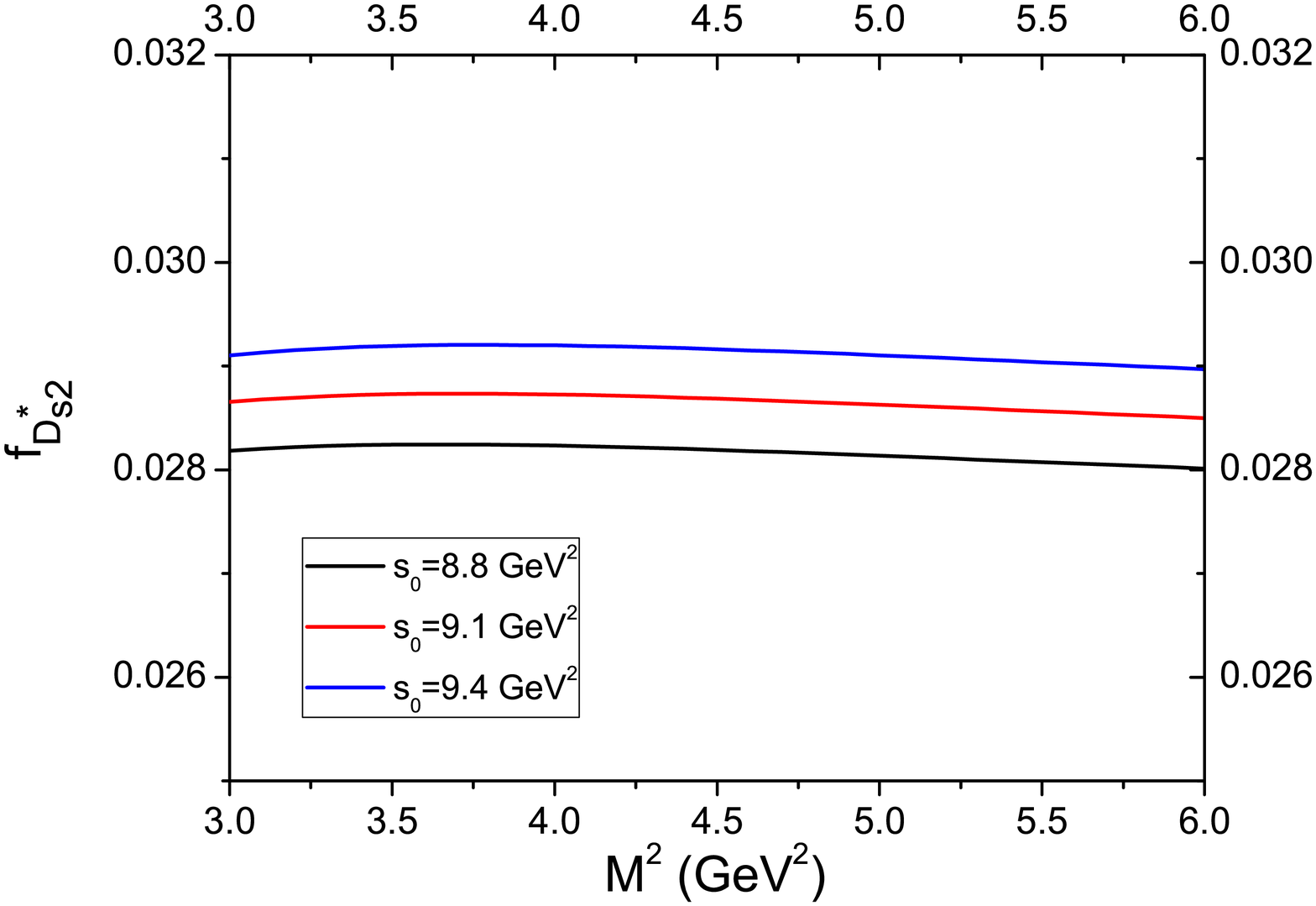}
\end{center}
\caption{Variations of the mass and decay constant
of the $D_{s2}^*(2573)$ meson with respect to temperature at $M^2$  at fixed values of the continuum threshold and at $T=0$.}
\label{Diagrams1}
\end{figure}
From these figures we see that the results are practically independent from the Borel mass parameter for the aforesaid Borel working region and the selected structure. Our numerical analysis show also that the
perturbative and non-perturbative parts overall constitute roughly $68\%$ and $32\%$ of the total ground state contribution, respectively. 
Moreover, the energy density constitutes about $10\%$ of the total non-perturbative contribution
in the working region of the Borel mass parameter. The rest contribution in the non-perturbative part comes from the quark condensate which also depend on the temperature according to Eq. (\ref{qbarq}).

 \begin{figure}[h!]
\begin{center}
\includegraphics[width=8cm]{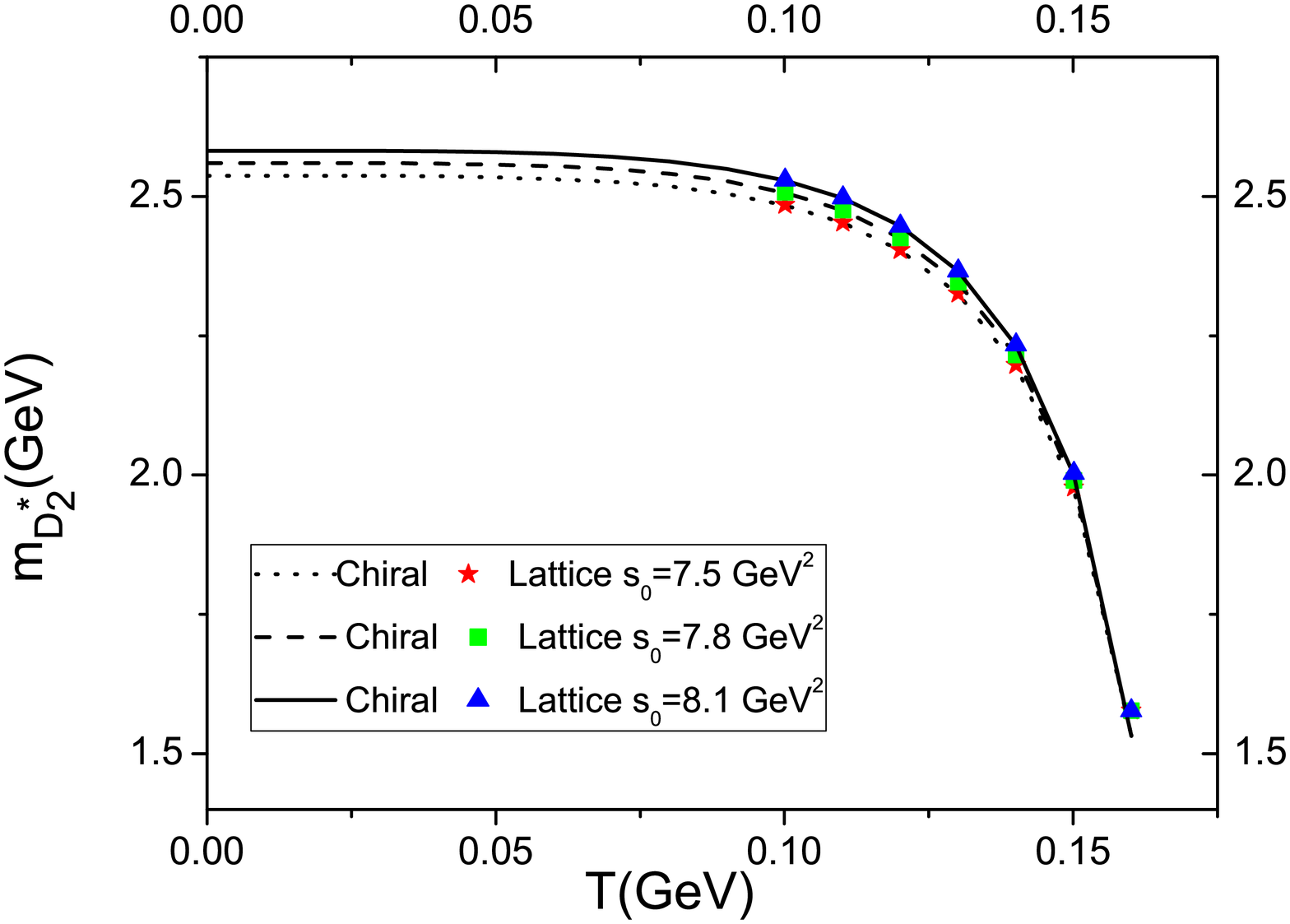}
\includegraphics[width=8cm]{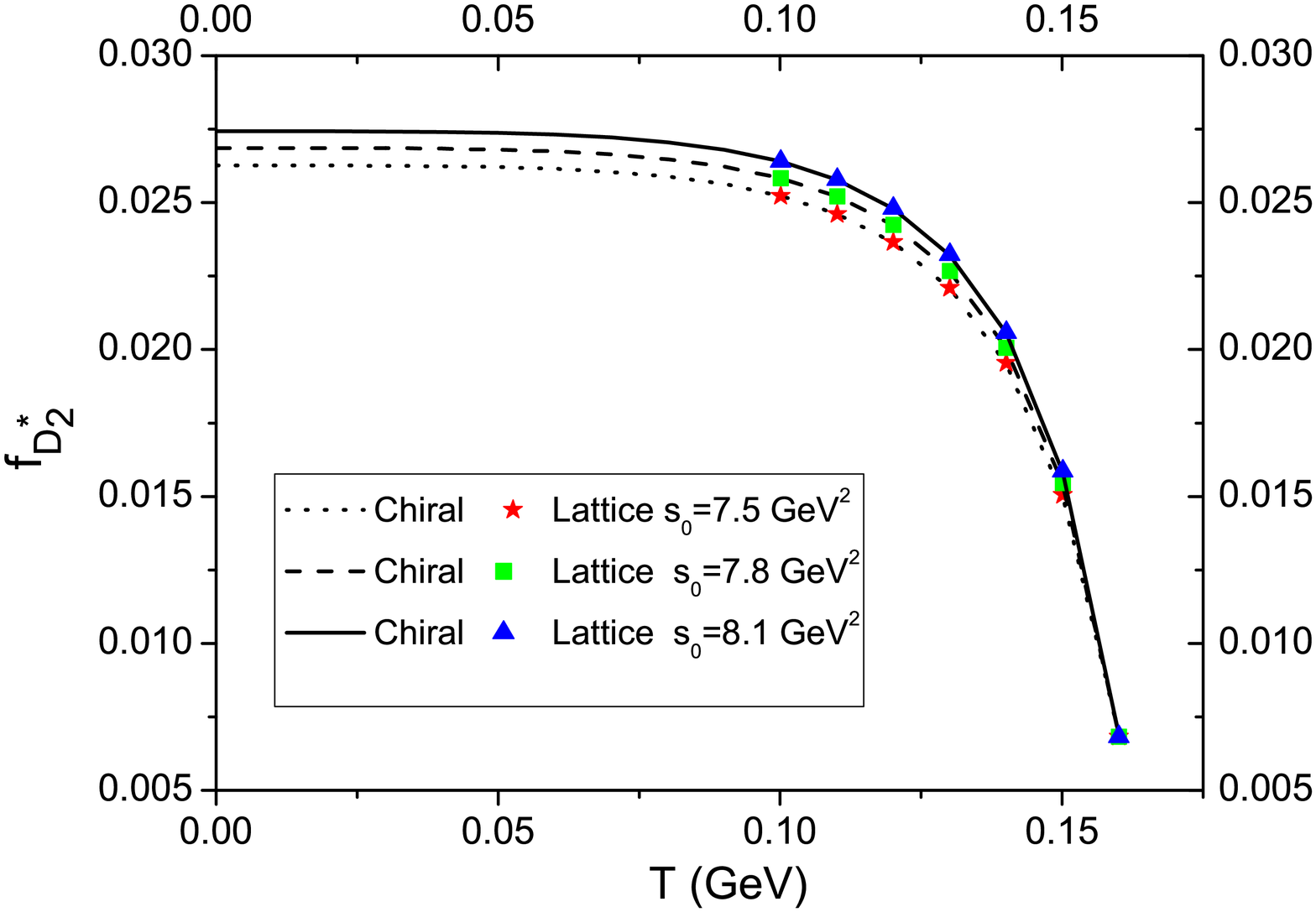}
\end{center}
\caption{Variations of the mass and decay constant
of the $D_2^*(2460)$ meson with respect to temperature at $M^2=3$ $GeV^2$.} \label{Diagrams1}
\end{figure}

\begin{figure}[h!]
\begin{center}
\includegraphics[width=8cm]{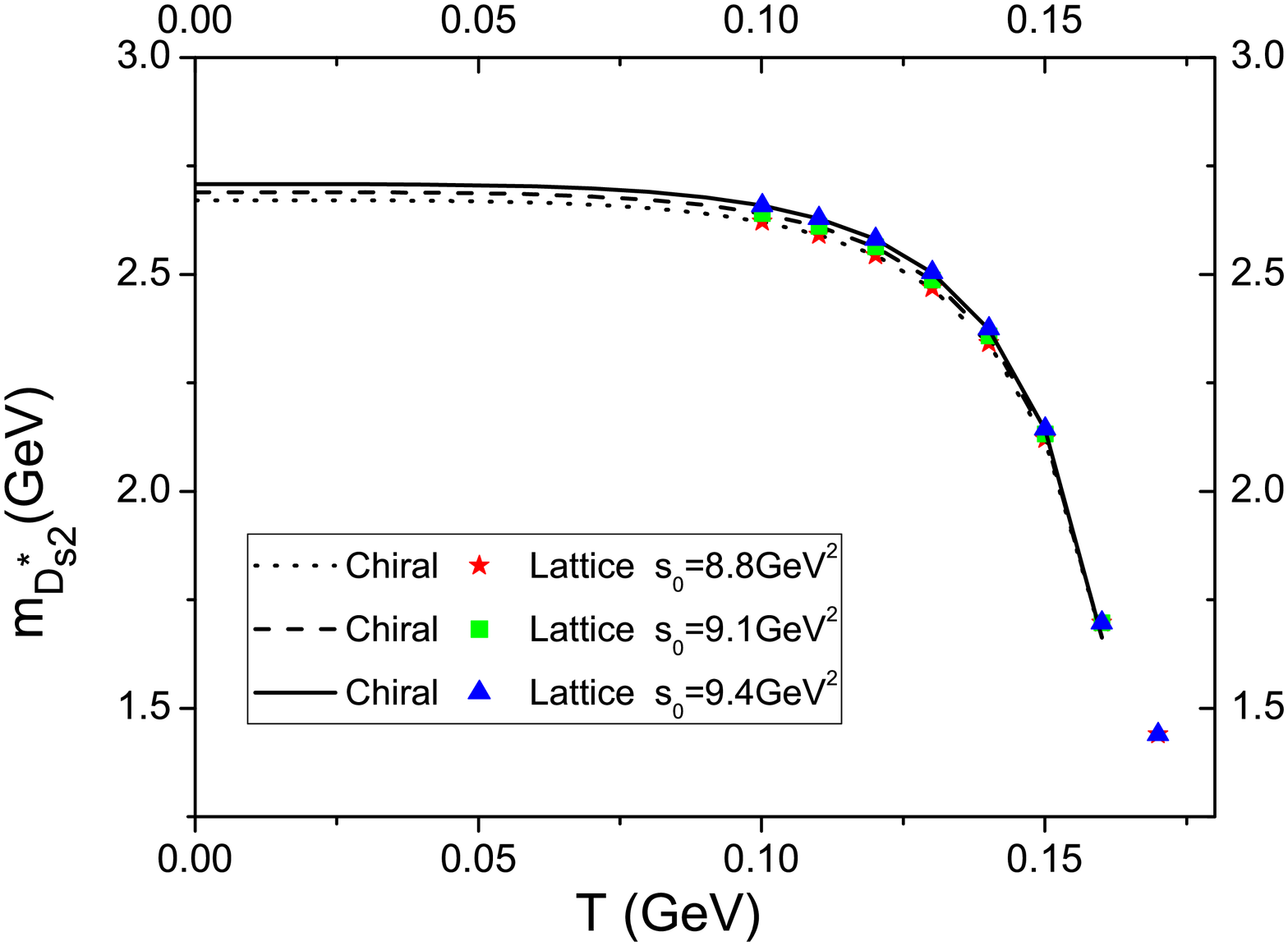}
\includegraphics[width=8cm]{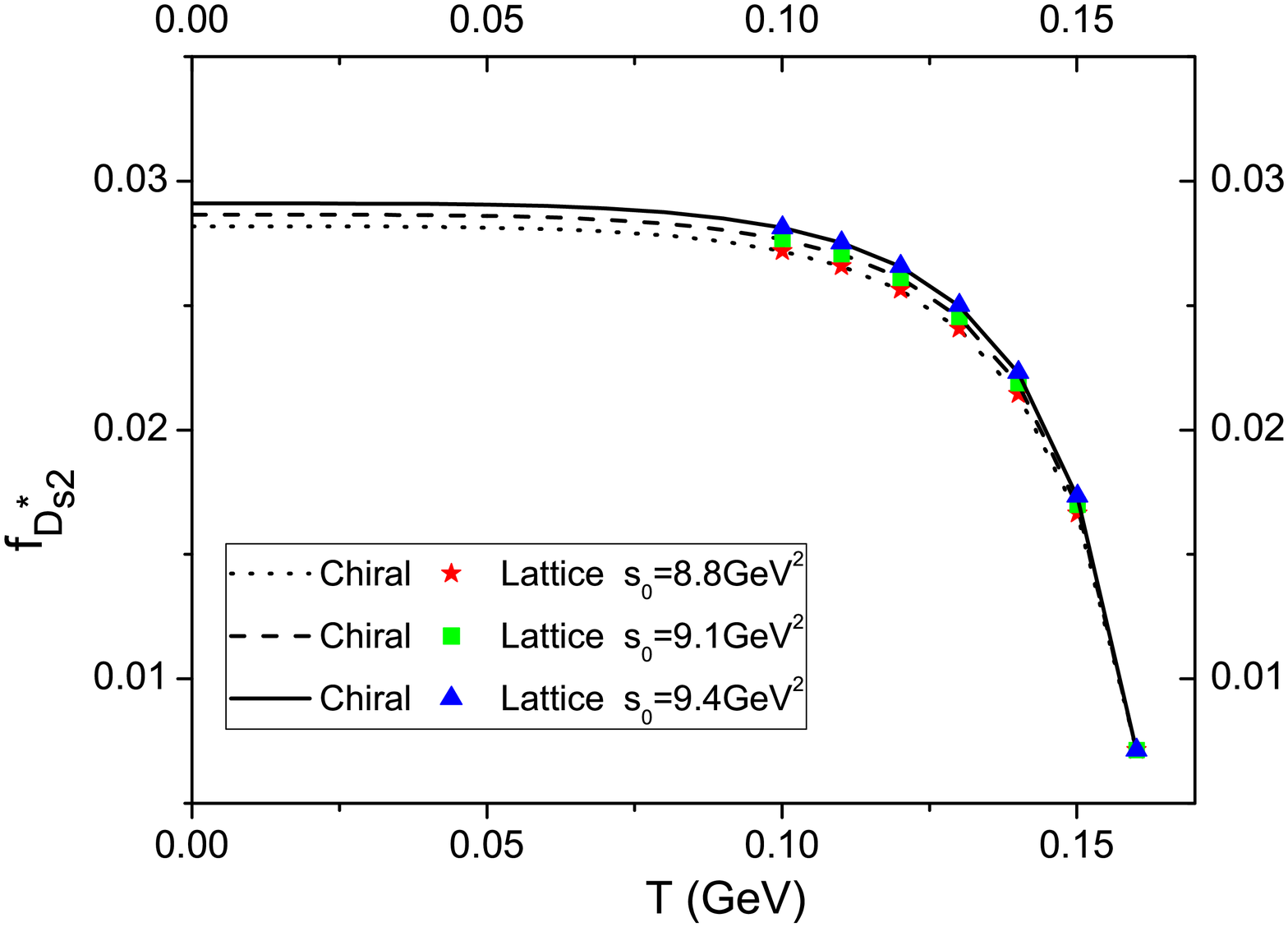}
\end{center}
\caption{Variations of the mass and decay constant
of the $D_{s2}^*(2573)$ meson with respect to temperature at $M^2=3$ $GeV^2$.}
\label{Diagrams1}
\end{figure}

Making use of all inputs we depict the variations of the masses and decay constants of the tensor mesons under consideration with respect to temperature in figures 4 and 5. From these figures
we read that the results remain approximately unchanged up to $0.1$ GeV, however, after this point, they start to diminish and fall considerably near to the critical temperature. These figures depict that 
the decay constants in both the Chiral and lattice parameterizations of the thermal average of the energy density reach roughly to $25\%$ of their values in vacuum,
while the masses decrease about $39\%$ and $37\%$ for $D_2^*$ and
$D_{s2}^*$ states, respectively. Our results at finite temperature  indicate also that the values of the physical quantities depend very weakly on the continuum threshold ($s_0$) such that near to the critical temperature
the results became practically independent of the continuum threshold at fixed value of the Borel mass parameter. It is also seen from these figures that the two Chiral and lattice parameterizations of the 
thermal average of the energy density lead to exactly the same results after $T=0.1~GeV$.

%
Our final task is to compare our results in the limit $T\rightarrow0$ with those previously obtained using vacuum sum rules as well as existing experimental data. This comparison is made in table 1. 
The errors quoted in this table for our results  are due to the uncertainties in determinations of the working regions for the 
continuum threshold and Borel mass parameter  as well as those
coming from the errors of other input parameters.
From this table we see
that, within the uncertainties, our predictions on the masses  of the tensor mesons  are 
consistent with the experimental data as well as the vacuum sum rules predictions \cite{hayriye,jale} with a good approximation. Our results on the decay constants 
are also roughly consistent with those of \cite{hayriye,jale} within the errors. 
 Our results on the leptonic decay constants can be checked in future experiments.

\begin{table}[h]
\renewcommand{\arraystretch}{1.5}
\addtolength{\arraycolsep}{3pt}
$$
\begin{array}{|c|c|c|c|c|c|}
\hline \hline
         &\mbox{Present\,\,\,Work} & \mbox{Experiment\,\,\cite{K.Nakamura}}& \mbox{Vacuum\,\,Sum\,\,Rules} \\
\hline
  \mbox{$m_{D_2^*(2460)}$(GeV)}        &  2.55\pm0.46    &  2.4626\pm0.0007   &2.53\pm0.45~~~~~~~~\cite{hayriye}\\
\hline
  \mbox{ $f_{D_2^*(2460)}$}        &  0.027\pm0.013 &  -  & 0.0228\pm0.0068 ~~\cite{hayriye}\\
\hline
  \mbox{$m_{D_{s2}^*(2573)}$(GeV)}        &  2.69\pm0.48    &  2.5719\pm 0.0008 &  2.55\pm0.44~~~~~\cite{jale}\\
\hline
  \mbox{ $f_{D_{s2}^*(2573)}$}        &  0.029\pm0.014 &  -  &  0.023\pm0.011  ~~~~~\cite{jale} \\

                    \hline \hline
\end{array}
$$
\caption{Values of the masses and decay constants of the
 tensor $D_2^*$ and $D_{s2}^*$ mesons at $T=0$.} \label{tab:lepdecconst}
\renewcommand{\arraystretch}{1}
\addtolength{\arraycolsep}{-1.0pt}
\end{table}

\section{Acknowledgment}
This work was supported in part by the Scientific and Technological
Research Council of Turkey (TUBITAK) under the research project
No. 110T284.

\end{document}